\documentclass[useAMS]{mn2e}
\usepackage{multirow}
\usepackage{epsfig}
\usepackage{subfigure}
\usepackage{url}
\usepackage[T1]{fontenc}
\usepackage{aecompl}

\def \nustar {{\it NuSTAR}}
\def \swift {{\it Swift}}
\def \swiftxrt {{\it Swift}/XRT}
\def \swiftbat {{\it Swift}/BAT}

\makeatletter
\let\jnl@style=\rm
\def\ref@jnl#1{{\jnl@style#1}}
\makeatother

\title[The hard X-ray spectrum of NGC5506]
{The hard X-ray spectrum of NGC 5506 as seen by $NuSTAR$}

\author[G. Matt et al]{G. Matt$^1$, M. Balokovi{\' c}$^2$, A. Marinucci$^1$, D.R. Ballantyne$^3$, 
S.E. Boggs$^4$, 
\newauthor F.E. Christensen$^5$, A. Comastri$^6$, W. W. Craig$^{5,7}$, P. Gandhi$^{8,9}$,  C.J. Hailey$^{10}$,
\newauthor F.A. Harrison$^2$, G. Madejski$^{11}$, K.K. Madsen$^{2}$, D. Stern$^{12}$ , W. W. Zhang$^{13}$\\
$^1$ Dipartimento di Matematica e Fisica, Universit\`a degli Studi Roma Tre, 
via della Vasca Navale 84, 00146 Roma, Italy\\
$^{2}$ Cahill Center for Astronomy and Astrophysics, California Institute of 
Technology, Pasadena, CA 91125, USA \\
$^3$ Center for Relativistic Astrophysics, School of Physics, Georgia Institute of Technology, 
Atlanta, GA, 30332, USA \\
$^4$ Space Sciences Laboratory, University of California, Berkeley, California 94720, USA \\
$^5$ DTU Space National Space Institute, Technical University of Denmark, Elektrovej 327, 
2800 Lyngby, Denmark\\
$^6$ INAF Osservatorio Astronomico di Bologna, via Ranzani 1, I-40127, Bologna, Italy\\
$^7$ Lawrence Livermore National Laboratory, Livermore, California 94550, USA \\
$^8$ Department of Physics, University of Durham, South Road, Durham DH1 3LE, UK \\
$^9$ School of Physics and Astronomy, University of Southampton, Highfield, Southampton, SO17 1BJ, UK \\
$^{10}$ Columbia Astrophysics Laboratory, Columbia University, New York, New York 10027, USA \\
$^{11}$ Kavli Institute for Particle Astrophysics and Cosmology, SLAC National Accelerator Laboratory, 
Menlo Park, CA 94025, USA \\
$^{12}$ Jet Propulsion Laboratory, California Institute of Technology, Pasadena, CA 91109, USA \\
$^{13}$ NASA Goddard Space Flight Center, Greenbelt, Maryland 20771, USA \\
}

\begin{document}
\maketitle
\label{firstpage}


\begin{abstract} 
{\it NuSTAR} observed the bright Compton-thin, narrow line Seyfert 1 galaxy, NGC~5506, 
for about 56 ks. In agreement with past observations,
the spectrum is well fit by a power law with $\Gamma\sim1.9$, a distant reflection component and 
narrow ionized iron lines. A relativistically blurred reflection
component is not required by the data. When an exponential high energy cutoff is added to the power law,
a value of 720$^{+130}_{-190}$ keV (90\% confidence level) is found. Even allowing  for 
systematic uncertainties, we find a 3$\sigma$ lower limit to the high-energy cutoff of 350 keV, the
highest lower limit to the cutoff energy found so far in an AGN by {\it NuSTAR}. 
\end{abstract}


\section{Introduction}

Thanks to its grazing incidence optics, {\it NuSTAR} (Harrison et al. 2013) is providing, for the first time,
source-dominated hard X-ray ($>$10 keV) observations of Active Galactic Nuclei
(AGN). The hard X-ray spectra can thus be studied in much greater detail than before,
and spectral parameters determined with unprecedented precision and robustness.

The intrinsic X-ray spectra of (radio-quiet) AGN is believed to be produced by Comptonization
of the accretion disk photons in a hot corona, with coronal temperatures well in excess of 10 keV 
(e.g. Perola et al. 2002, Malizia et al. 2014).
One of the main goals of the {\it NuSTAR} AGN program is to determine the 
coronal parameters (temperature, optical depth, location, geometry) and, at the very least, the exponential 
high-energy cutoff which, together with the power law index, encodes information about these 
parameters.

Precise measurements of the high-energy cutoff (Brenneman et al. 2014a,b, Marinucci
et al. 2014a, Balokovi{\' c} et al. 2014, Ballantyne et al. 2014), or interesting lower limits 
to it (Matt et al. 
2014, Marinucci et al. 2014b) have already been obtained by {\it NuSTAR} for several AGN. 
In this paper we study 
the high-energy spectrum of the bright, nearby ($z$=0.006181) Compton-thin (Wang et al. 1999)
narrow line Seyfert 1 Galaxy (Nagar et al. 2002), NGC~5506. 

NGC~5506 has been observed by all major X-ray satellites. In the first XMM-{\it Newton} observation, 
clear evidence of narrow, neutral, and ionized iron lines
were found (Matt et. al 2001, Bianchi \& Matt 2002, 
Bianchi et al. 2003), but no evidence has ben found of a broad component. One was later
found by Guainazzi et al. (2010), albeit rather weak, when analysing all eight XMM-{\it Newton}
observations obtained between February 2001 and January 2009
(see also Patrick et al. 2012 for {\it Suzaku} results). Simultaneous XMM-{\it Newton} and {\it BeppoSAX}
observations permitted the search for a high energy cutoff, which was found to be 
at 140$^{+40}_{-30}$ keV,
but with large systematic uncertainties due to ambiguities in the modeling (Bianchi et al. 2003, 2004).

This paper is structured as follows: Sec.~2 describes the observation and data reduction;
Sec.~3 presents the spectral analysis, while the results are discussed in Sec.~4.

\begin{figure*}
\includegraphics[scale=.52]{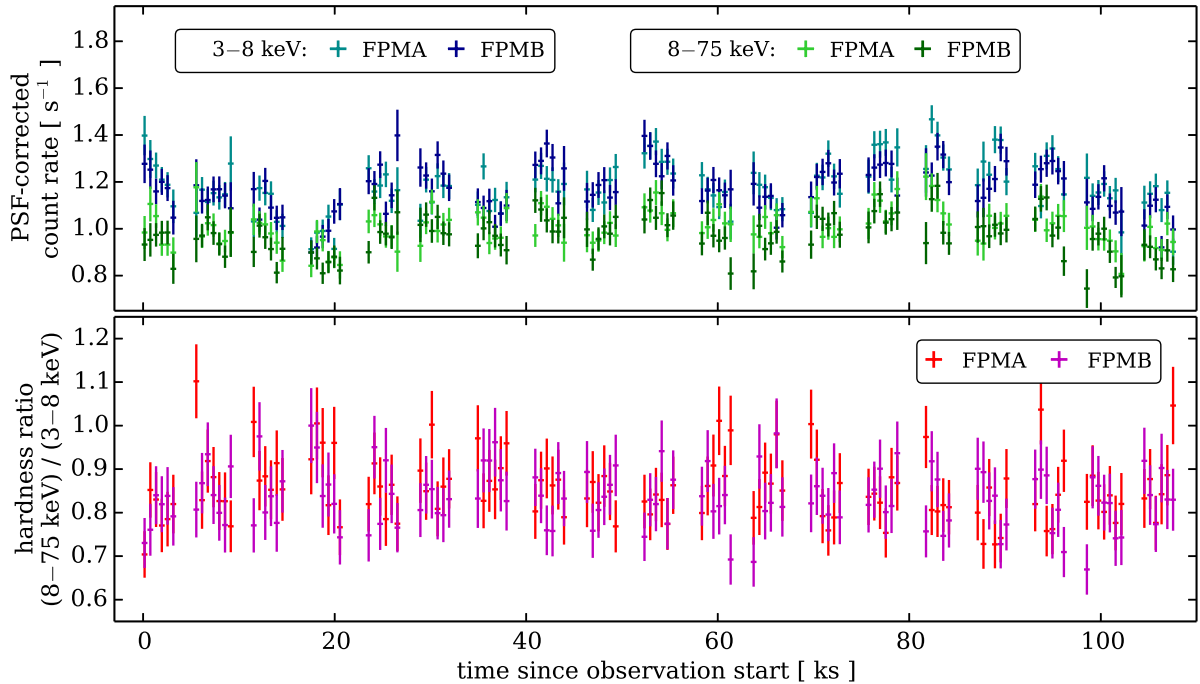}
\caption{Upper panel: FPMA and FPMB light curves in the 3-8 keV and 8-75 keV energy ranges.
Bins of 600 s are used. Lower panel: the (8-75 keV)/(3-8 keV) ratio.}
\label{lc}
\end{figure*}

\section{Observations and Data reduction}

NGC~5506 was observed with \nustar\ (OBSID 60061323) on 2014~April~1, as part of a snapshot 
survey of AGN selected from the \swiftbat\ all-sky survey (e.g. Baumgartner et al., 2013). 
The observation was coordinated with the \swift\ observatory (OBSID 00080413001), which 
observed the same target starting just 12 hours later, on 2012~April~2. The \nustar\ and 
\swift\ observations are sufficiently close in time that they can be considered simultaneous
as there is no evidence for spectral variability on this time scale (see Fig.~\ref{lc}). The
combined data  
therefore provide an instantaneous broadband snapshot of the X-ray spectrum of NGC~5506 in the 
0.8--80~keV band. The total \nustar\ and \swiftxrt\ exposures are 56~ks and 2.6~ks, respectively.

The \nustar\ data were reduced in the standard manner described in detail in Perri et al. (2013). 
We used HEASOFT~v\,6.15.1, NuSTARDAS~v\,1.3.1, and CALDB version 20131223. Following the event 
filtering using the {\tt nupipeline} script, we extracted the source spectrum from a circular 
aperture 80\arcsec ~in radius centered on the peak of the point source image. The background 
extraction region covered the free area of the same detector, excluding the $\simeq$110\arcsec 
~region around the source in order to avoid flux in the Point Spread Function wings. The spectrum and the 
corresponding response files were generated using the {\tt nuproducts} script. We bin the
spectra for modules FPMA and FPMB to a minimum of 20~counts per bin before background 
subtraction, and fit them simultaneously, without coadding. The cross-normalization constant 
is left free to vary in all our fits (with the instrumental normalization of FPMA fixed at unity).

We used online resources provided by the ASI Science Data Center (ASDC) for \swiftxrt\ 
data reduction,\footnote{http://www.asdc.asi.it/mmia/} HEASOFT~v\,6.13 and CALDB 
version 130313 were used for processing. The spectrum was extracted from a region with a radius of 
20\arcsec ~centered on the brightest peak of emission, and the background was sampled from an annular 
region extending between 40\arcsec ~and 80\arcsec ~around the source. For spectral fitting we 
use the source spectrum binned to a minimum of 20~counts per bin before background subtraction. 
The cross-normalization constant is left free to vary in all our fits. We find it to be consistent 
with the expected systematic uncertainty of $\simeq10$\% determined from cross-calibration between
 \nustar\ and \swiftxrt\ (Madsen et al., in prep.).

\section{Spectral analysis}

The analysis of the {\it NuSTAR} light curves shows the presence of flux variability of the order of 20\%,
but without any significant spectral variability (Fig.~\ref{lc}). A hint of a hardening
of the spectrum at lower fluxes is apparent, but it is easily explained by a small
contribution from reflection from distant matter. Therefore, we used the spectra integrated
over the entire observation. 
In all spectral fitting we combined the {\it NuSTAR} and \swiftxrt\ spectra, and included Galactic
absorption with a column density of 3.8$\times$10$^{20}$ cm$^{-2}$.

All spectral fitting was done using the public package {\sc xspec}. The reader
is referred to its webpage\footnote{http://heasarc.gsfc.nasa.gov/docs/xanadu/xspec/} 
for informations on the fitting code and for 
details on the various spectral models applied here. Unless otherwise stated, 
all quoted errors correspond to 90\% confidence levels for one interesting parameter. 

\begin{figure}
\includegraphics[angle=-90,scale=.32]{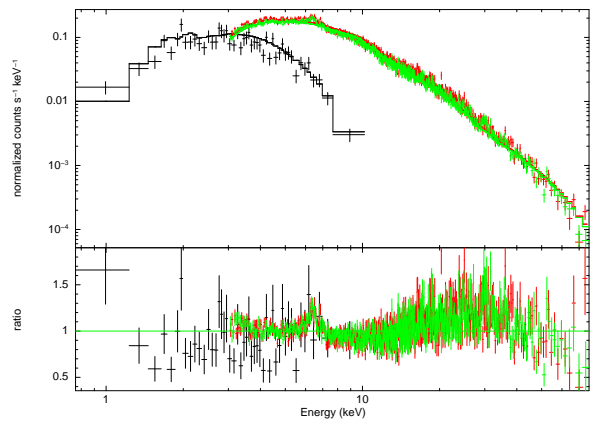}
\caption{Spectra, best fit model, and data/model ratio when fitting with an absorbed, cutoff
power law. Black refers to \swiftxrt\ data, red to {\it NuSTAR} FPMA and green to  
{\it NuSTAR} FPMB.
The background is also plottes, showing that the source is larger than the background
in the whole band.}
\label{badfit}
\end{figure}

\begin{figure}
\includegraphics[angle=-90,scale=.32]{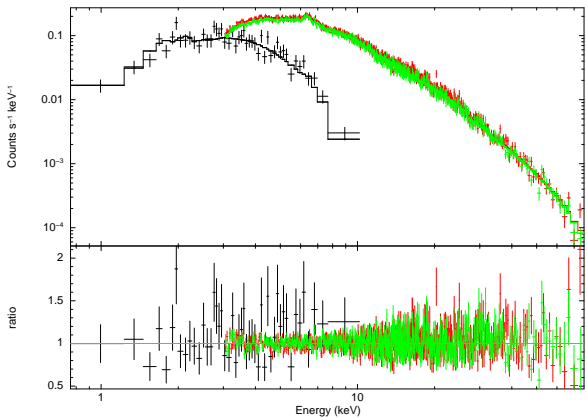}
\caption{ Spectra, best fit model, and data/model ratio for the model composed
of an absorbed, cutoff power law, a reflection component and ionized iron lines 
(see Table~\ref{tablebestfit}). Colors as in Fig.~\ref{badfit}. }
\label{bestfit}
\end{figure}

\begin{figure}
\includegraphics[angle=-90,scale=.32]{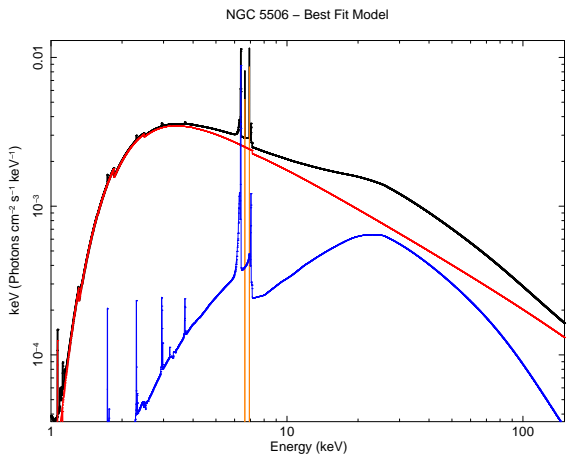}
\caption{Best fit model for the model composed of: an
absorbed cutoff power law (in red), a reflection component (in blue), 
ionized iron lines (in orange). The top black line is the total spectrum.}
\label{model}
\end{figure}

First, we fitted the spectra with a simple, absorbed power law with an exponential
high-energy cutoff ({\sc cutoffpl} model in {\sc xspec}).
The fit is unacceptable ($\chi^2$/d.o.f.=1989/1090), mostly due to a prominent iron line and
curvature at high energies (see Fig.~\ref{badfit}). 
As both features are suggestive of reflection, we added
an {\sc xillver} component (actually the {\sc xillver-a-Ec2} version, which includes both
the angular dependence of the emitted radiation and a primary power law with an 
exponential cutoff\footnote{http://hea-www.cfa.harvard.edu/$\sim$javier/xillver/}).  
This results in a dramatic improvement in the quality
of the fit ($\chi^2$/d.o.f.=1150/1085). 
Further improvements are found by adding a {\sc mekal}
component (absorbed only by Galactic material)
to account for some residuals in the soft X-ray band, which are likely
due to the extended emission found by {\it Chandra} (Bianchi et al. 2003), and which result in
$\chi^2$/d.o.f.=1144/1083; and adding 
Fe {\sc xxv} and Fe {\sc xxvi} narrow lines, following Matt et al. (2001)
($\chi^2$/d.o.f.=1135/1081). The best fit parameters are listed in Table~\ref{tablebestfit},
while the spectra, best fit model and data/model ratio are shown in Fig~\ref{bestfit}.
 The various spectral components are shown in Fig.~\ref{model}).
The normalization of the reflection component corresponds to a standard $R$ parameter (defined
as the solid angle of the reflecting matter in units of 2$\pi$) of about 0.7.  
The observed 2-10 keV flux is 4.87$\times$10$^{-11}$ erg cm$^{-2}$ s$^{-1}$; 
when corrected for absorption,
the flux is 6.23$\times$10$^{-11}$ erg cm$^{-2}$ s$^{-1}$ corresponding to a luminosity of 
5.26$\times$10$^{42}$ erg s$^{-1}$. This flux is at the lower end of the range of fluxes found so
far for this source (Guainazzi et al. 2010).

\begin{figure}
\includegraphics[scale=.38]{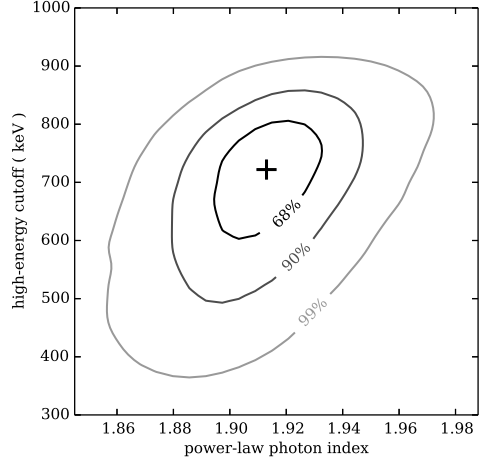}
\caption{ Power law photon index and high-energy cutoff contour plot, 
calculated for the model
composed of an absorbed, cutoff power law, a reflection component and ionized iron lines.
}
\label{contour}
\end{figure}

\begin{figure}
\includegraphics[scale=.38]{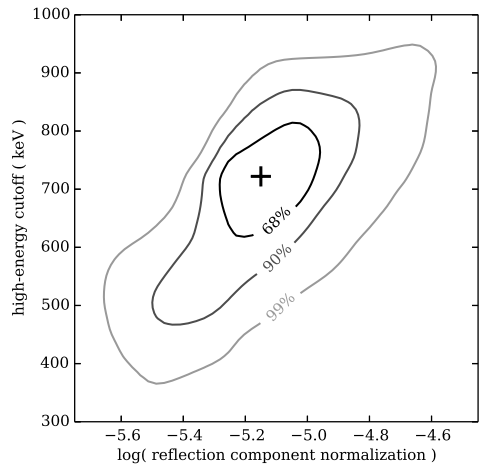}
\caption{Normalization of the reflection component and high-energy cutoff contour plot, 
calculated for the model
composed of an absorbed, cutoff power law, a reflection component and ionized iron lines.
}
\label{contour2}
\end{figure}

Undoubtedly, the most interesting result relates to the high-energy cutoff. Figs.~\ref{contour}
and~\ref{contour2},
which show the correlation between the high-energy cutoff vs. the power law
 index and the normalization of the reflection component, respectively, 
indicate that the cutoff
is very high (best fit value around 700 keV). Even with this very large value, the high-energy
cutoff is constrained on both its lower and upper ends at the 3$\sigma$ level. This shows the amazing
capability of {\it NuSTAR} to measure this parameter (at least for a bright source with a relatively
simple spectrum like NGC5506), even for energies much higher than the 79 keV upper limit to the
{\it NuSTAR} energy range. 
Even allowing for some remaining systematic errors in the effective area, 
which may affect the upper bounds of the contours in Figs.~\ref{contour}
and~\ref{contour2}, and given
the fact (see below) that the result is robust against different modelling of the continuum,
we can affirm that the cutoff energy is, at the 99.97\% confidence level, higher than 350 keV
(see the Appendix for further details).

The reflection component appears to be moderately ionized. Indeed, forcing the reflecting
matter to be neutral results in a fit which is worse at the 99.99\% confidence level, according
to an F-test. The measurement of the cutoff energy, however, remains basically unchanged.
The iron abundance is 1.6 times higher than (but almost consistent with) solar,
while the inclination of the reflecting surface with respect to the 
line of sight is found to be less than 44$^{\circ}$. In an obscured source, in  
which the inclination angle of the accretion disk and the broad line region 
is supposed to be high, this may indicate that the surface of the 
reflecting matter (which in the fitting model is assumed to be a slab) is misaligned with 
the disk surface, as indeed expected if it has a toroidal geometry.

Regarding the iron lines, we note that the flux of the neutral component
(if fitted separately from the reflection component) is 6.87$(^{+0.71}_{-1.05})$$\times$$10^{-5}$ 
ph cm$^{-2}$ s$^{-1}$, consistent with most XMM-{\it Newton} observations but somewhat 
larger than the last two ones (Guainazzi et al. 2010). 
The fluxes of the He- and H-like lines
are also consistent with most XMM-{\it Newton} observations, but a bit lower in case of
two observations (Guainazzi et al. 2010). 
They are consistent with Suzaku measurements (Patrick et al. 2012).

Guainazzi et al. (2010) suggested the presence of a weak, broad component of the iron line.
Therefore, we added an {\sc xillver-A-Ec2} component blurred by relativistic effects ({\sc kdblur}
model). The $\chi^2$/d.o.f.=1127/1076 tells us that this component is required only at the 82\% 
confidence level, according to an F-test. That the reflection is mostly due to distant matter
is corroborated by the fact that the flux of the neutral iron line (when fitted separately
with a narrow gaussian) is consistent with the values found in past observations, despite the
large variations in the flux of the continuum.
Not surprisingly, given the weakness of the relativistically
blurred reflection the system parameters (among them the black hole spin) 
are unconstrained, and all other parameters -- 
including the high-energy cutoff -- remain basically unchanged.

\begin{table}
\begin{center}
\caption{Best fit parameters for the model composed by 
an absorbed, cutoff power law, a reflection component and 
ionized iron lines. See text for details.}
\begin{tabular}{cc}
\hline\hline
N$_{\rm H}$ (10$^{22}$ cm$^{-2}$) & 3.10$^{+0.21}_{-0.20}$ \\
$\Gamma$ & 1.91$\pm$0.03  \\
$E_c$ (keV) &  720$^{+130}_{-190}$  \\
$\xi$ (erg cm s$^{-1}$) & 22$^{+15}_{-7}$ \\
$A_{Fe}$ & 1.63$\pm$0.58 \\
$i$ & $<$44$^{\circ}$ \\
I$_{\rm Fe XXV}$ (ph cm$^{-2}$ s$^{-1}$)  &  $<$14.1$\times$10$^{-6}$ \\
I$_{\rm Fe XXVI}$ (ph cm$^{-2}$ s$^{-1}$) & 8.4($\pm$6.4)$\times$10$^{-6}$ \\ 
\hline
\label{tablebestfit}
\end{tabular}
\end{center}
\end{table}

Finally, assuming as customary that the primary power law spectrum is due to Comptonization of thermal
disk photons by hot electrons in a 
corona, we tried to estimate the coronal parameters by substituting the cutoff power law
with a Comptonization model, namely {\sc comptt}. For simplicity, 
the parameters of the reflection component,
apart from the normalization, were kept frozen to the best fit values found with the cutoff
power law. A slab geometry and seed photon temperature of 20 eV have been adopted. 
The fit is good ($\chi^2$/d.o.f.=1135/1084). The optical depth is found to be 
0.02$^{+0.19}_{-0.01}$, and the coronal electron temperature 440$^{+230}_{-250}$ keV  (consistent, within
the errors, to the standard 2-3 factor between temperature and cutoff energy). 
Similar results are obtained with a spherical geometry, 
apart from a larger optical depth (about 0.09), as
expected (in the spherical geometry the optical depth is the radial, thence effective, one while
in the slab geometry the vertical - thence lower than effective - optical depth is used).
Using instead the  {\sc compps} model, a temperature of about 270 keV and an optical depth
of  0.06 (0.14) are found for the slab (sphere) geometry.

\section{Discussion}

We report on a joint {\it NuSTAR} and {\it Swift} observations of the bright,
obscured Narrow Line Seyfert 1 galaxy NGC~5506. The hard X-ray spectrum is composed of an absorbed
(column density of about 3$\times$10$^{22}$ cm$^{-2}$) power law (with $\Gamma\sim$1.9) with an 
exponential high-energy cutoff, 
plus a moderately ionized reflection component, and ionized iron lines. 
The presence of relativistically blurred reflection is not required by the data.

The most interesting result is the measurement of the high-energy cutoff, which demonstrates
the capability of {\it NuSTAR} to constrain this. The best fit value is 
720$^{+130}_{-190}$ keV (90\% confidence
level for one interesting parameter). Even at the 3$\sigma$ level, the cutoff value is formally 
bound on both
sides (see Fig.~\ref{contour}). Even allowing for possible systematic errors, we can conservatively
put a firm lower limit to the cutoff energy of about 350 keV at the 99.97\% confidence level.

This is the highest lower limit to the cutoff energy found so far in an AGN by {\it NuSTAR}, and it is 
definitely inconsistent with values found in other AGN studied by {\it NuSTAR},
e.g. in SWIFT~J2127.4+5654 (108$\pm$11 keV, Marinucci
 et al. 2014a), MCG-5-23-16 (116$\pm$6 keV, Balokovi{\' c} et al. 2014),
IC~4329A (186$\pm$14 keV, Brenneman et al. 2014), and 3C~382 ($214^{+147}_{-63}$ keV in one of 
the two {\it NuSTAR} pointings, Ballantyne et al. 2014)
Other lower limits, like those found in Ark~120 ($>$190 keV, Matt et al. 2014), 3C382 ($>$190 
keV in the other {\it NuSTAR} pointing) and in 
NGC~2110 ($>$210 keV, Marinucci et al. 2014b), while interesting, are nevertheless 
lower than the one found here in NGC~5506.

The above results, which suggest that very high cutoff energies do exist but are not too
common, are still in agreement with the predictions of X-ray background (XRB) synthesis models. 
Gilli et al. (2007) showed that even a mean value of 300 keV
would already saturate the XRB at 100~keV. Therefore, sources like NGC~5506 should be
the exception rather than the rule.

While the number of AGN with a precise and robust measurement of the high-energy cutoff 
is still too small to search for correlations with system parameters, we cannot help noting
that this parameter is quite different in the two best studied narrow line Seyfert 1 
galaxies observed so far by {\it NuSTAR}, namely NGC~5506 and
SWIFT~J2127.4+5654 (even if we must note that NGC 5506 has X-ray properties, like e.g. the 
power law index and the variability pattern, more similar to broad than that to narrow lines Seyfert 
1s). The black hole mass of NGC~5506 is unfortunately very poorly known, with 
estimates ranging from 5$\times$10$^6$ to 10$^8$ solar masses (see Guainazzi et al. 2010 
and references therein), and the Eddington ratio is correspondingly uncertain, ranging
from 0.007 to 0.14 (assuming that the bolometric luminosity is 20 times the 2-10 keV one). 
If the low value (indicated by the black hole mass measurement based on the central velocity 
dispersion) is the correct one, the two strictest lower limits to the 
cutoff energy, NGC~5506 and NGC~2110, are both found in sources with relatively low accretion rates, 
perhaps indicating inefficient cooling of the corona due to the low UV/soft X-ray flux.

\section*{Acknowledgements}

This work has made use of data from the {\it NuSTAR} mission,
a project led by the California Institute of Technology,
managed by the Jet Propulsion Laboratory, and funded by the
National Aeronautics and Space Administration. We thank
the {\it NuSTAR} Operations, Software and Calibration teams for
support with the execution and analysis of these observations.
This research has made use of the {\it NuSTAR} Data Analysis Software (NuSTARDAS) jointly
developed by the ASI Science Data Center (ASDC, Italy) and the California Institute of
Technology (USA). 
GM, AM and AC acknowledge financial support from Italian Space Agency under grant 
ASI/INAF I/037/12/0-011/13, GM and AM also from the European
Union Seventh Framework Programme (FP7/2007-2013) under grant agreement
n.312789. MB acknowledges support from the Fulbright
International Science and Technology Award.


\section*{Appendix. How robust is the cutoff energy measurement?}

As the main result of this paper is the tight lower limit to the cutoff energy,
we tested the robustness of the results by making a few changes in either
the data reduction or in the spectral modelling.

We adopted different extraction regions (40\arcsec, 60\arcsec, 80\arcsec, 100\arcsec and 120\arcsec), and
different binnings (20 counts per bin, 50 counts per bin, 100 counts per bin, S/N$>$5),
 and fitted the data with
all possible combinations. The best fit cutoff energy varies somewhat, but remains always
higher than 400 keV (and in most cases higher than 450-500 keV) at the 90\% confidence level
apart from the 40\arcsec aperture (and 20 counts per bin) where the lower limit is 350 keV, very likely
due to the lower number of source counts. The use of the Cash statistics
(Cash 1979) instead of $\chi^2$ statistics also does not significantly affect the results. 

Another test we made was to substitute the {\sc xillver} with the {\sc reflionx} (Ross \& Fabian 
2005) reflection model, in a version which allows the primary emission to be a power law with 
an exponential cutoff. The latter model has no angular dependence, being an average 
over the emission angle. The fit is worse ($\chi^2$/d.o.f.=1150/1082, to be compared 
with $\chi^2$/d.o.f.=1135/1081 which we found using {\sc xillver}), and a cutoff energy of 
$\sim$300 keV is found, with similar values for the other parameters. 
In this case, however, the addition of a relativistically blurred component
results in a more significant fit improvement ($\chi^2$/d.o.f.=1124/1077, significant 
at the 99.98\% confidence level according to the F-test), and the cutoff energy is $>$390 keV.
The ionization parameter and iron abundance are consistent with those found with the
{\sc xillver} model, while the power law index is somewhat steeper, $\sim$2. The best fit
emissivity index, inner disk radius and inclination angle are 1.84$^{+1.05}_{-0.14}$, $<$2.6 gravitational 
radii and 84$\pm$4 degrees, respectively, while
the outer disk radius has been kept fixed to 400 gravitational radii.) 

Finally, we note that the relative accuracy of the {\it NuSTAR} effective area at
70 keV is about 5\% (Madsen et al., in prep.). This means that the error on the high-energy cutoff is likely
dominated by statistical, rather than systematic, uncertainties. Indeed, adding a 5\%
systematic uncertainties to all energy bins - clearly a gross overestimate - only slightly
larger errors on $E_c$ are found: the best fit value is, 
adopting the same model as in Table~\ref{tablebestfit},
756$^{+139}_{-233}$ keV at 90\% confidence level, and 756$^{+232}_{-325}$ keV at 3$\sigma$.

\end{document}